\documentclass[aps, reprint, amsmath,amssymb, pra, superscriptaddress]{revtex4-2}

\usepackage{color}
\usepackage{graphicx}
\usepackage{dcolumn}
\usepackage{bm}
\usepackage{physics}
\usepackage{hyperref}
\usepackage{url}
\usepackage{subcaption}
\usepackage{amsmath}
\usepackage{notes2bib}
\usepackage{ dsfont }
\usepackage{svg}
\usepackage{verbatim}
\usepackage{amssymb}
\usepackage{amsfonts}              
\usepackage{amsthm}                
\usepackage{mathtools}
\usepackage{kotex}
\usepackage{qcircuit}
\usepackage[normalem]{ulem}
\usepackage{xcolor}

\newtheorem{theorem}{Theorem}
\newtheorem{prop}[theorem]{Proposition}

\newtheorem{corollary}[theorem]{Corollary}
\newtheorem{lem}[theorem]{Lemma}

\bibnotesetup{note-name    = ,use-sort-key = false} 
\begin{document}

\title{
Randomness for quantum channels:\\Genericity of catalysis and quantum advantage of uniformness}

\author{Seok Hyung Lie}
\author{Hyunseok Jeong}
\affiliation{%
 Department of Physics and Astronomy, Seoul National University, Seoul, 151-742, Korea
}%

\date{\today}

\begin{abstract}
Randomness can help one to implement quantum maps that cannot be realized in a deterministic fashion. Recently, it was discovered that explicitly treating a randomness source as a quantum system could double the efficiency as a catalyst for some tasks. In this work, we first show that every quantum channel that can be implemented with a randomness source without leaking information to it must be a catalysis. For that purpose, we prove a new no-go theorem that generalizes the no-hiding theorem, the no-secret theorem that states no quantum information can be shared with other system as a secret without leaking some information. Second, we show that non-degenerate catalysts should be used classically when no extra dimension is allowed, which leads to the fact that the quantum advantage of a catalytic process strictly comes from the uniformness of the randomness source. Finally, we discuss a method to circumvent the previous result that achieves quantum advantage with non-degenerate catalyst uniformized by employing extra work space.
\end{abstract}

\pacs{Valid PACS appear here}
\maketitle

\section{Introduction}

Randomness is a universal resource for numerous applications. Its usage ranges from everyday tasks such as shuffling playing cards to information processing tasks such as symmetric-key cryptography \cite{delfs2007symmetric} and randomized computation \cite{smid2000closest}. Recently, the role of randomness as a catalyst for the quantum state transition and the information masking process has been studied \cite{boes2018catalytic, boes2019neumann, lie2019unconditionally, lie2020randomness}. The catalycity of randomness means that the randomness is not depleted during the process. Remarkably, it was discovered that, for some tasks, the efficiency of a uniform randomness source can be doubled when the source is explicitly treated as a quantum system, compared to the case where the source is treated as a classical randomness source such as coin tossing or dice roll \cite{boes2018catalytic, lie2020randomness}.

On the other hand, the resource theory of quantum randomness is still in its initial stage, and many important questions are left unanswered. Is the catalycity of randomness limited only to some specific cases? Can an arbitrary type of randomness be used as a catalyst if its entropic measures are sufficiently high? What is the origin of the advantage of quantum randomness source?

To answer these questions, in this work, we advance the theory of quantum randomness for arbitrary randomness sources. To distinguish the role as a randomness source from the role as an information dump of ancillary systems in quantum information theory, we define the concept of randomness-utilizing process in which no information flows to ancillary system while implementing a quantum channel.

\begin{figure}
    
    \includegraphics[width=.25\textwidth]{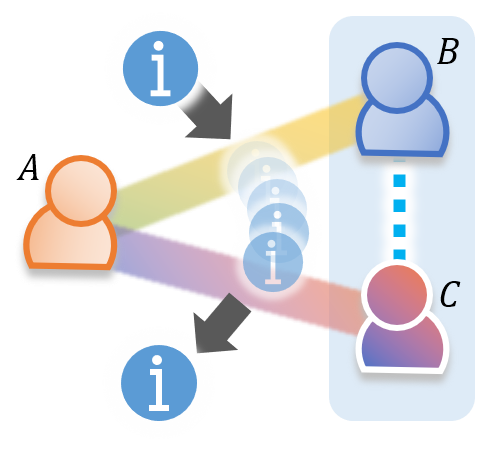}
    \centering
    \caption{Assume that $A$ implements a quantum channel by using $B$ as an ancillary system without leaking information to $B$, where systems $BC$ are initially prepared in a pure state. The no-secret theorem states that systems $AC$ can always recover the input state of the channel. No quantum information can be shared with other system as a secret without leaking some information.}
    \label{fig:nosecret}

\end{figure}

Next, we prove a new no-go result that we call \textit{the no-secret theorem} which generalizes the no-hiding theorem \cite{braunstein2007quantum} and the no-masking theorem \cite{modi2018masking} stating that no quantum information of a quantum system, however partial it is, cannot be shared with other system as a secret without leaking some information to it. Based on the no-secret theorem, we show that catalycity, the conservation of randomness source throughout the process, is a generic phenomenon by proving that every dimension-preserving randomness-utilizing processes is a catalysis. Even dimension non-preserving processes are catalytic if two different processes that transform the randomness source in converse ways are used alternatively.

Second, we prove that \textit{uniformness} is the source of the advantage of catalytic quantum randomness. To this end, we first show that there exists a gap between the upper bounds of achievable efficiencies of classical and quantum randomness sources therefore quantum advantage is universal for randomness-utilizing processes. It is then demonstrated that non-degenerate randomness sources can be used only as a classical catalyst. In light of the fact that non-degeneracy is generic for probability distributions, it follows that additional efforts such as uniformization are required in order to take advantage of quantum randomness.

Finally, despite the newly found restrictions, adopting an operationally natural generalization of randomness-utilizing processes, we obtain a resource theory of randomness where randomness is depletable and catalycity is nontrivial. In this more general setting, in return for requiring more work space, any randomness source with sufficiently large entropy can be used as catalytic quantum randomness regardless of its degeneracy.

This paper is organized as follows. In section \ref{subsec:cata}, we prove the no-secret theorem and show that catalysis is generic among randomness-utilizing processes. In section \ref{subsec:degen}, we show that the advantage of quantum randomness source comes from the degeneracy, or the uniformness, of a randomness source. In section \ref{subsec:nonuni}, we introduce a method that can circumvent the restriction and utilize a nonuniform randomness source. In section \ref{subsec:conc}, we summarize the paper and discuss open problems.

\section{Main results}
\subsection{Genericity of catalysis} \label{subsec:cata}
Every quantum channel can be realized with unitary interaction with an ancillary system, according to the Stinespring dilation theorem \cite{stinespring1955positive}. Considering that no quantum information can be destroyed by unitary evolution, for every irreversible quantum channel, a role of the ancillary system is storing information removed from the main system. It is demonstrated in the extreme case by the no-hiding theorem \cite{braunstein2007quantum} (and equivalently the no-masking theorem \cite{modi2018masking}), which states that when a quantum state is disappeared from a system, then it should be recoverable from its purification system, i.e. environment. Therefore, implementation of quantum channel seemingly leaks information to the ancillary system, which is true for initially pure ancillary state because of the conservation law of quantum information \cite{lie2020randomness}. 

On the other hand, the space of quantum correlation of mixed bipartite state is very vast and capable of containing the whole space of local quantum state, which was shown by the possibility of $((2,2))$-threshold secret sharing or randomized quantum masking \cite{cleve1999share,lie2019unconditionally,lie2020randomness}. It means that one can implement an erasure map, which completely destroys the information of an input state, by utilizing the correlation between two systems, not the local marginal state of ancillary system itself, as its information dump. In that situation, even though the information itself is not destroyed and could be faithfully recovered globally, still no local system can access to the information. Does it mean that the erased information is the secret between and only between them?

The answer is negative, since every purification of $((2,2))$-threshold quantum secret sharing scheme is a $((2,3))$-threshold quantum secret sharing scheme \cite{cleve1999share,gottesman2000theory}, meaning that quantum state shared as a secret with the ancillary system can be also restored with its purification system. In short, no quantum state can be shared as secret between only two systems. One can ask if this result holds for general quantum channels other than erasure channels. Maybe this result is the consequence of trying to hide the whole quantum state, in contrast to hiding partial information such as classical information within quantum system. To answer this question, we first give a formal definition of implementation of quantum channel without leaking information to its local ancillary system.

We denote quantum systems by uppercase alphabets $(A,B,\dots)$ and their corresponding Hilbert spaces as $\mathcal{H}_A$. The space of operators on $\mathcal{H}$ will be written as $\mathcal{B}(\mathcal{H})$. We will say a map defined on $\mathcal{B}(\mathcal{H})$ is $d$-dimensional if $\dim\mathcal{H}=d$. In this work, we will only consider finite-dimensional systems. For an ancillary system not to gain information through the implementation of quantum channel, it should not depend on the input state of the channel. In that case, we can say that the ancillary system only functions as a source of randomness. Therefore, we say that a quantum channel $\Phi$ on $\mathcal{B}(\mathcal{H}_A)$ is \textit{randomness-utilizing} when it can be expressed as
 \begin{equation} \label{eqn:rand}
     \Phi(\rho) = \Tr_B U (\rho \otimes \sigma) U^\dag,
 \end{equation}
 with some unitary operator $U$ on $\mathcal{H}_A \otimes \mathcal{H}_B$ and a \textit{randomness source} $\sigma$, which is a quantum state on $\mathcal{H}_B$, and $\Tr_A U (\rho \otimes \sigma) U^\dag$ is a constant quantum state independent of $\rho$. We will sometimes call the whole process $U (\rho \otimes \sigma) U^\dag$, not the channel $\Phi$ itself, a randomness-utilizing process.   The second condition is imposed since we only want the randomness source to provide randomness to the given process and do not want it to function as an information storage. In fact, if we do not impose the second condition, any quantum map can be expressed in the form of (\ref{eqn:rand}) by using Stinespring dilation. We will call the constant output of $\Tr_A U (\rho \otimes \sigma) U^\dag$ corresponding to a randomness-utilizing quantum process as the \textit{residue randomness} of the process.
 
 When the residue randomness has the same spectrum (the set of eigenvalues including degeneracy) with the randomness source, we say the randomness-utilizing process is \textit{catalytic} or the process uses the randomness catalytically. A catalytic channels  is a channel that has a catalytic randomness-utilizing process implementation. It is because, in that case, one can use the residue randomness as the randomness source of the same process for another uncorrelated input.
 
 In the following we will use the family of R\'{e}nyi entropies $\{S_\alpha\}$ given as \cite{renyi1961measures}
 \begin{equation}
     S_\alpha(\rho)=\frac{1}{1-\alpha}\log\Tr\rho^\alpha,
 \end{equation}
 for $0< \alpha$, where the $\log$ is the logarithmic function with base 2. We also define the max-entropy $S_0(\rho):=\lim_{\alpha\to0} S_\alpha(\rho)=\log \rank\rho$ and the min-entropy $S_\infty(\rho):=\lim_{\alpha\to\infty}(\rho)=-\log\max_i\rho_i$ where $\{\rho_i\}$ is the spectrum of $\rho$. Note that $S_1:=\lim_{\alpha \to 1} S_\alpha$ is the usual von Neumann entropy.
 
 Now we are ready to prove the following result, which we call \textit{the no-secret theorem}. Here, we say that a bipartite unitary $W_{XY}$ restores the input state $\rho$ of the system $X$ of channel $\Psi(\rho)$ that maps $\rho$ to a bipartite state of the system $XY$ if $Tr_YW_{XY}\Psi(\rho)W_{XY}^\dag=\rho$ for every $\rho$.
 
 \begin{theorem} [The no-secret theorem] \label{thm:nosecret}
     Assume that $\sigma_B$ is a quantum state whose purification is $\ket{\Sigma}_{BC}$ on the system $BC$. For any randomness-utilizing quantum channel $\Phi$ acting on $A$ implemented with $\sigma_B$ as the randomness source, the input state of $\Phi$ can be restored with a unitary operator on $AC$. 
 \end{theorem}
 \begin{proof}
     Assume that $\tau_B$ is the residue randomness of the process and $\ket{T}_{BC}$ is its purification.
     Following the notation of Eqn. (\ref{eqn:rand}), for a maximally entangled state $\ket{\Gamma}_{RA}:=\frac{1}{\sqrt{d}}\sum_{i=1}^d \ket{i}_R\ket{i}_A$, the definition of randomness-utilizing process can be equivalently expressed as the following equation through the Choi-Jamio{\l}kowski isomorphism \cite{choi1975completely,jamiolkowski1972linear},
     \begin{equation} \label{eqn:puri}
         \Tr_A U_{AB}(\dyad{\Gamma}_{RA}\otimes\sigma_B)U_{AB}^\dag=\frac{1}{d}\mathds{1}_R\otimes\tau_B.
     \end{equation}
     A purification of the left hand side is $U_{AB}\ket{\Gamma}_{RA}\otimes\ket{\Sigma}_{BC}$, and a purification of the right hand side is $\ket{\Gamma}_{RA}\otimes\ket{T}_{BC}$. Since every purification of the same mixed state is unitarily similar to each other on the purification system, there exists a unitary operator $V_{AC}$ on the system $AC$ such that
     \begin{equation} \label{eqn:puri2}
         U_{AB}\ket{\Gamma}_{RA}\otimes\ket{\Sigma}_{BC}=V_{AC}\ket{\Gamma}_{RA}\otimes\ket{T}_{BC}.
     \end{equation}
     It follows that $\Tr_{BC} V_{AC}^\dag U_{AB}(\rho_A\otimes\sigma_B)U_{AB}^\dag V_{AC}=\rho_A$, which implies that the input state $\rho$ is restored by applying the unitary operator $V_{AC}^\dag$ on $AC$.
 \end{proof}
 
  The no-secret theorem says that it is impossible to share \textit{any} quantum information with some party, not limited to sharing the whole quantum state, without leaking some information. For example, in quantum masking with pure states \cite{modi2018masking}, hiding phase information of a quantum system in a bipartite state is possible, but it accompanies the leakage of amplitude information.
  
 Actually, the no-secret theorem is a stronger no-go result than the no-hiding theorem (or equivalently the no-masking theorem) since a stronger version of the no-hiding theorem can be derived from the no-secret theorem. Here, an irreversible quantum channel $\mathcal{C}$ is a channel that has no recovery channel $\mathcal{R}$ such that $\mathcal{R}\circ\mathcal{C}(\rho)=\rho$ for any input state $\rho.$ An erasure channel is one example of irreversible channel.
 
 \begin{corollary}[Stronger no-hiding theorem]
     No irreversible quantum channel can be implemented without leaking some information to the ancillary system initially prepared in a pure state.
 \end{corollary}
 \begin{proof}
     We follow the notations of the proof of Theorem \ref{thm:nosecret}, but we assume that $\sigma_B$ is a pure state this time, i.e. $\sigma_B=\dyad{s}_B$ , hence its purification should be a product state $\ket{\Sigma}_{BC}=\ket{s}_B\ket{t}_C$. We negate the stronger no-hiding theorem and assume that an irreversible $\Phi$ can be implemented through a randomness-utilizing process with a unitary operator $U_{AB}$ and a pure randomness source. The system $C$ in a pure state $\ket{t}_C$, however, need to be uncorrelated to any other system, so the marginal state of $AC$ should be in the product state $\Phi(\rho)_A\otimes\dyad{t}_C$ for any input state $\rho_A$. From the no-secret theorem, there exists a unitary operator $V_{AC}^\dag$ acting on $AC$ that recovers the input state $\rho,$ i.e. $Tr_C V_{AC}^\dag \Phi(\rho)_A\otimes \dyad{t}_C V_{AC}=\rho_A$. However, it implies that the quantum channel $\mathcal{R}(\cdot):=Tr_C V_{AC}^\dag \Phi(\cdot)_A\otimes \dyad{t}_C V_{AC}$ is the recovery map of $\Phi$, which contradicts the assumption that $\Phi$ is an irreversible quantum channel.
 \end{proof}
 
 From the proof of Theorem \ref{thm:nosecret}, one can see that both $U_{AB}$ and $V_{AC}$ implement the same quantum channel on the system $A$ from their identical Choi matrices, but the transformation of their randomness sources are converse to each other. Hence the following Corollary is obtained.
 
 \begin{corollary} \label{coro:reverse}
     For any randomness-utilizing process that transform the source of randomness as $\sigma \to \tau$, there exists another randomness-utilizing implementation of the same quantum channel that transforms the source of randomness as $\tau \to \sigma$.
 \end{corollary}
 
 Randomness-utilizing process usually randomizes its input states, and by doing so it decays information. There are the two most typical examples of such processes, dephasing and erasure maps. By dephasing map with respect to a basis $\{\ket{i}\}$ we mean quantum maps of the form
 $$\mathcal{D}(\rho)=\sum_i \bra{i}\rho\ket{i}\dyad{i}.$$
 Similarly by erasure map, we mean quantum maps of the form
 $$\mathcal{E}(\rho)=\tau,$$
 with some fixed quantum state $\tau$. However, if we try to implement an erasure map as a randomness-utilizing process, then it is proven that \cite{imai2005information, lie2019unconditionally, lie2020randomness} the output state $\tau$ should have the von Neumann entropy larger than $\log d$, where $d$ is the dimension of the input state's Hilbert space. Therefore if we insist the output system of the erasure map has the same dimension as the input system, then the output state of the map must be the maximally mixed state, i.e. $\frac{\mathds{1}}{d}$. Afterwards, by the erasure map, we mean the constant quantum map that outputs the maximally mixed state, which is also known as the completely depolarizing map.
 
 In Ref. \cite{boes2018catalytic}, a special case of randomness-utilizing dephasing map was studied, where the randomness source is limited to be maximally mixed state, i.e.  a uniform randomness source and the whole process is required to be catalytic. The lower bound of the size of the randomness source was derived in Ref. \cite{boes2018catalytic} with this restriction, which is half the size of the system being dephased. One might ask, however, if this randomness non-consuming property is a special property that other generic randomness-utilizing processes do not have. First, we show that randomness-utilizing implementation of dimension-preserving quantum channels should never decrease the amount of randomness.
 
 Here, that a probability distribution $p=(p_i)_{i=1}^n$ majorizes another distribution $q=(q_i)_{i=1}^n$, i.e. $p \succ q$, means that $\sum_{i=1}^k p_i \geq \sum_{i=1}^k q_i$ for all $k=1,\dots,n$ and for quantum states $\rho \succ \sigma$ means that their spectra are in majorization relation. A dimension-preserving quantum map is a quantum map whose input and output systems have the same finite dimension, so that their Hilbert spaces are isomorphic.
 
 \begin{prop} \label{prop:major}
    For any dimension-preserving randomness-utilizing quantum channel transforming its randomness source as $\sigma \to \tau$, the initial randomness majorizes the residue randomness, i.e. $\sigma \succ \tau$.
 \end{prop}
 \begin{proof}
     Consider an arbitrary randomness-utilizing quantum channel $\mathcal{C}: \mathcal{B(H}_A) \to \mathcal{B(H}_A)$ and its randomness source $\sigma$ with unitary operator $W$ on $\mathcal{H}_A \otimes \mathcal{H}_B$ such that
    \begin{equation}
        \mathcal{C}(\rho)=\Tr_B W(\rho \otimes \sigma)W^\dag,
    \end{equation}
    and $\Tr_A W(\rho \otimes \sigma)W^\dag=\tau$  for any state $\rho$. Now we define $\eta_{AB}:=W(\frac{\mathds{1}}{d} \otimes \sigma)W^\dag$. Then we evaluate the $\alpha$-R\'{e}nyi entropy of $\eta_{AB}$, i.e. $S_\alpha(\eta_{AB})$, which is same as $S_\alpha(\frac{\mathds{1}}{d}\otimes \sigma)= \log d + S_\alpha(\sigma)$, because of the fact that unitary operators do not change the R\'{e}nyi entropy and the additivity of the R\'{e}nyi entropy.
    Next, from the weak subadditivity of the R\'{e}nyi entropy \cite{van2002renyi}, i.e.
    \begin{equation} \label{eqn:weak}
        S_\alpha(\eta_{AB})\leq S_0(\eta_A) + S_\alpha(\eta_B),
    \end{equation}
    we have $\log d + S_\alpha (\sigma) \leq S_0(\mathcal{C}(\frac{\mathds{1}}{d})) + S_\alpha(\tau) \leq \log d + S_\alpha(\tau)$ since $S_0(\eta_A) \leq \log d$ as $A$ is a $d$-dimensional quantum system. Thus we get $S_\alpha (\sigma) \leq S_\alpha (\tau)$ for any $\alpha \geq 0$. It implies $\sigma \succ \tau$.
 \end{proof}
 This result provides an important perspective on the randomness consumption of quantum processes: it is not randomness \textit{per se} that is consumed in the process, but it is its \textit{uncorrelatedness} with other system, which is often referred to as \textit{privacy}.
 
 Combined with Corollary \ref{coro:reverse}, we obtain the following Theorem that says the catalytic usage of quantum randomness is generic.
 
 \begin{theorem} \label{thm:dimcat}
     Every dimension-preserving randomness-utilizing process is catalytic.
 \end{theorem}
 \begin{proof}
     If a dimension-preserving randomness-utilizing process transforms its randomness source as $\sigma \to \tau$, by Corollary \ref{coro:reverse}, there must be another dimension-preserving randomness-utilizing process that transforms its randomness source as $\tau \to \sigma$. From Proposition \ref{prop:major}, we get both $\sigma \succ \tau$ and $\tau \succ \sigma$, which is possible only when their spectra are identical, which in turn implies that the whole process is catalytic.
 \end{proof}
 
We also obtained a significant constraint on the set of quantum channels that can be implemented through randomness-utilizing process. Here, a unital channel $\Phi$ is a quantum channel that preserves the identity operator, i.e. $\Phi(\mathds{1})=\mathds{1}$.

\begin{theorem} \label{thm:unital}
    Only unital quantum channels among dimension-preserving channels can be implemented through randomness-utilizing process.
\end{theorem}
\begin{proof}
    We use the assumptions and notations of the proof of Proposition \ref{prop:major}. This time, we use the subadditivity of von Neumann entropy \cite{araki2002entropy} for $\eta_{AB}=W(\frac{\mathds{1}}{d}\otimes\sigma)W^\dag$, i.e.
    \begin{equation}
        S(\eta_{AB})\leq S(\eta_{A})+S(\eta_B).
    \end{equation}
    Here, $S(\eta_{AB})=S(\frac{\mathds{1}}{d}\otimes\sigma)=\log d + S(\sigma)$ and $S(\eta_B)=S(\sigma)$ as $\eta_B=\sigma$ from the catalycity. It follows that $\log d \leq S(\eta_A)$, which is achievable only when $\eta_A=\mathcal{C}(\frac{\mathds{1}}{d})=\frac{\mathds{1}}{d}$, i.e. $\mathcal{C}$ is unital.
\end{proof}

Since every unital channel never decreases entropy \cite{frank2013monotonicity}, Theorem \ref{thm:unital} implies that every (dimension-preserving) randomness-utilizing channel not only can be implemented with a randomness source but also only can randomize its input states.

\begin{figure}
    
    \includegraphics[width=.25\textwidth]{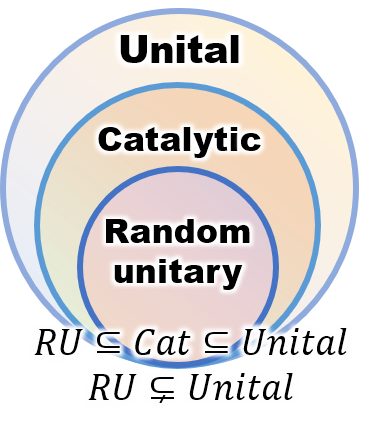}
    \centering
    \caption{Inclusion relations between the sets of random unitary ($RU$), catalytic ($Cat$) and unital ($Unital$) channels. It is known that in general dimension $RU$ and $Unital$ are not identical. It is still unknown if the inclusions $RU\subseteq Cat$ and $Cat \subseteq Unital$ are proper.}
    \label{fig:inc}

\end{figure}

From Theorem \ref{thm:dimcat} and \ref{thm:unital}, we can see that the set of catalytic channels forms an interesting subclass of the set of unital channels that contains the set of random unitary channels (See  FIG. \ref{fig:inc}.). The von Neumann-Birkhoff theorem \cite{birkhoff1946three} states that every doubly stochastic matrix can be expressed as a convex sum of permutations. However, it is known that the quantum counterpart of doubly stochastic matrix, unital map, does not allow an expression in the form of convex sum of unitary operations \cite{landau1993birkhoff}. In other words, the von Neumann-Birkhoff theorem does not hold in quantum mechanics. It implies that the set of random unitary channels is a proper subset of the set of unital channels. We still do not know if every unital channel is catalytic or every catalytic channel is a random unitary channel. 

We can observe that the set of catalytic channels is another natural quantum generalization of the set of permutation operations in the sense that both operations being mixed and the usage of randomness are quantum, in contrast to he classical usage of randomness in random unitary channels. Therefore we conjecture a quantum version of von Neumann-Birkhoff theorem: Every unital channel is a catalytic channel. At this point, we only know that all three sets are convex from the following Proposition.

\begin{prop}
    The set of catalytic channels is convex.
\end{prop}
\begin{proof}
    Let $\Phi_1$ and $\Phi_2$ be catalytic channels on the same system $A$ that have respective catalytic processes given as $\Phi_1(\rho)=\Tr_{B_1} U_1(\rho\otimes \sigma_1)U_1^\dag$ and $\Phi_2(\rho)=\Tr_{B_2} U_2(\rho\otimes \sigma_2)U_2^\dag$. Note that systems $B_1$ and $B_2$ can be systems with different dimensions. Then, for any $0\leq p \leq 1$, any convex combination $\Phi=p\Phi_1 + (1-p)\Phi_2$ can be catalytically implemented with catalyst $\sigma=\sigma_0\otimes\sigma_1\otimes\sigma_2$ on system $B=B_0B_1B_2$ where $\sigma_0=p\dyad{0}_{B_0}+(1-p)\dyad{1}_{B_0}$ is 2-dimensional mixed state with the controlled unitary $U=\dyad{0}_{B_0}\otimes U_1 \otimes \mathds{1}_{B_2}+\dyad{1}_{B_0}\otimes \mathds{1}_{B_1} \otimes U_2$, i.e. $\Phi(\rho)=\Tr_{B} U(\rho \otimes \sigma)U^\dag$.
\end{proof}

Corollary \ref{coro:reverse} also has a very significant consequence for dimension non-preserving randomness-utilizing processes. As there are two ways to implement the same randomness-utilizing map that maps the randomness source in both directions, e.g. $\sigma \to \tau$ and $\tau \to \sigma$, it follows that every randomness-utilizing channels can be implemented catalytically when two processes are used \textit{alternatively}. It shows that indeed catalysis is generic among randomness-utilizing processes.

\begin{theorem} \label{thm:couple}
    For arbitrary randomness-utilizing quantum channel $\Phi$ on $A$, there is a catalytic randomness-utilizing process that implements $\Psi$ on two copies of $A$, i.e. $A_1A_2$ such that $\Tr_{A_1}\Psi(\rho_{A_1}\otimes\sigma_{A_2})=\Phi(\sigma)$ and $\Tr_{A_2}\Psi(\rho_{A_1}\otimes\sigma_{A_2})=\Phi(\rho)$ for all $\rho$ and $\sigma$.
\end{theorem}
 We remark that Theorem \ref{thm:couple} has a striking formal resemblance with the result of Ref. \cite{mendl2009unital}, which states that $O(d)$-covariant unital channels that are not random unitary operations, a special class of catalytic processes, can become one by taking two copies of it. However, also note that $\Psi$ in Theorem \ref{thm:couple} is different from a simple two-copy version of $\Phi,$ i.e. $\Phi^{\otimes2}$, since two parties can be correlated even for product inputs.

\subsection{Quantum advantage of degeneracy} \label{subsec:degen}

 Next, we investigate the nature of catalytic quantum randomness.  To do so, we first examine the previously assumed conditions on randomness sources. In this section, we assume that every randomness-utilizing channel is dimension-preserving. In Ref. \cite{gour2015resource}, \textit{noisy operations} were considered, which are the quantum maps of the form of (\ref{eqn:rand}) but with uniform randomness sources. In the resource theory of nonequilibrium, maximally mixed states are considered free since it can be interpreted that they have reached equilibrium, so that they are useless in the thermodynamic setting. In Ref. \cite{boes2018catalytic}, however, the same noisy operation formalism is adopted for resource-theoretic approach to randomness. From that perspective, maximally mixed state is no longer free but a highly desirable form of randomness compared to nonuniform randomness \cite{mcinnes1990impossibility, dodis2004possibility}.

However, randomness sources are in general nonuniform and usually require some kind of uniformization for applications \cite{von195113}. A canonical example of such randomness source is thermal state with non-degenerate Hamiltonian. In fact, almost every finite probability distribution and quantum state is non-degenerate and any degenerate probability distribution can be turned into a non-degenerate one with arbitrarily small perturbation. The following theorem shows that almost every randomness source cannot be used quantumly.

\begin{theorem} \label{thm:nondeg}
    Any randomness-utilizing process using non-degenerate randomness source as a catalyst is a random unitary map in which randomness is used classically.
\end{theorem}

\begin{proof}
    We continue from the proof Proposition \ref{prop:major}, but we can assume that now $\mathcal{C}$ is an arbitrary randomness-utilizing unital map by Theorem \ref{thm:dimcat}. As initial and residue randomness are unitarily similar, i.e. $\tau=R\sigma R^\dag$ for some unitary operator $R$, by making $W$ absorb $R$, without loss generality we can assume $\tau=\sigma$. Let us define the `reciprocal' channel of $\mathcal{C}$ for each input $\rho$,
    \begin{equation}
        \hat{\mathcal{C}_\rho}(\xi):=\Tr_A W(\rho \otimes \xi)W^\dag.
    \end{equation}
     Observe that $\sigma$ is a fixed point of $\hat{\mathcal{C}_\rho}$ for arbitrary $\rho$. Consider the case of $\rho=\frac{\mathds{1}}{d}$. For this case, $\hat{\mathcal{C}_\frac{\mathds{1}}{d}}$ is an unital quantum channel and one can decompose $\hat{\mathcal{C}_\frac{\mathds{1}}{d}}$ into Kraus operators $\{K_{nm}\}$ such that $\hat{\mathcal{C}_\frac{\mathds{1}}{d}}(\xi)=\frac{1}{d}\sum_{nm} K_{nm} \xi K_{nm}^\dag $ given as $K_{nm}=(\bra{n} \otimes \mathds{1}) W (\ket{m} \otimes \mathds{1})$. Since $\mathcal{H}_A$ is a finite-dimensional Hilbert space, $\sigma$ being a fixed point of $\hat{\mathcal{C}_\frac{\mathds{1}}{d}}$ implies that every $K_{nm}$ commutes with $\sigma$ \cite{arias2002fixed}. However, since $\sigma$ is assumed to be non-degenerate, it implies that every $K_{nm}$ is diagonal in the eigenbasis of $\sigma$. As a result the bipartite unitary $W$ is diagonal in the system $B$, i.e. $W$ is a controlled unitary of the form
    \begin{equation}
        W=\sum_m W_m^A \otimes \dyad{m}_B,
    \end{equation}
    where $W_m$ are unitary operators on $\mathcal{H}_A$ and $\sigma=\sum_m q_m \dyad{m}$ is the unique spectral decomposition of $\sigma$. Therefore we get the following random unitary expression of the channel $\mathcal{C}$,
    \begin{equation}
        \mathcal{C}(\rho)=\sum_m q_m W_m \rho W_m^\dag.
    \end{equation}
    It implies that the usage of randomness in this process is classical, i.e. $\mathcal{C}$ is implemented by applying $W_m$ depending on the random variable $m$ sampled from the distribution $\{q_m\}$. 
\end{proof}

When we say a probability distribution $(p_i)$ is used classically, we mean that it is used to implement the convex sum of deterministic processes, i.e. unitary maps, in the form of random unitary like $\sum_i p_i U_i \rho U_i^\dag$. Note that even if we give up the exact implementation of the desired map, the requirement of catalycity still forces the approximate map to be a random unitary map. Being forced to use randomness classically undermines the efficiency of randomness-utilizing process. 

Hereby we examine the quantum advantage of randomness usage in resource theory of randomness for non-degenerate randomness sources. The following Theorem unifies the pre-existing results on the advantage of using quantum randomness sources. Here, the entanglement-assisted classical capacity of a quantum channel $\mathcal{N}$,  $C_{EA}(\mathcal{N})$, is the classical capacity achievable with the channel $\mathcal{N}$ with pre-distributed entangled state between two parties.

\begin{theorem} \label{thm:bound}
    A $d$-dimensional randomness-utilizing unital channel with the entanglement-assisted classical capacity $C_{EA}$ requires a classical randomness source with at least $ 2\log d - C_{EA}$ of min-entropy or a quantum randomness source with at least $ \log d -\frac{1}{2}C_{EA}$ of min-entropy.
\end{theorem}
\begin{proof}

Theorem \ref{thm:bound} follows from Theorem 2 of Ref. \cite{lie2020randomness}.  We state it here for the completeness.

\begin{lem} \label{lem:cap}
    Consider a quantum channel $\mathcal{N}$, a convex sum of quantum channels $\{\mathcal{N}_i\}$, i.e. $\sum_i p_i \mathcal{N}_i = \mathcal{N}$. For all $i$, the difference of the entanglement-assisted classical capacity $C_{EA}$ of $\mathcal{N}_i$ and $\mathcal{N}$ has the following upper bound,
    \begin{equation} \label{ineq1}
        C_{EA}(\mathcal{N}_i) - C_{EA}(\mathcal{N}) \leq -\log p_i.
    \end{equation}
\end{lem}

Every  randomness-utilizing process $\Phi(\rho)=\Tr_B U(\rho \otimes \sigma)U^\dag$ can be expressed as a convex sum of the form $\Phi(\rho)=\sum_i p_i \Phi_i(\rho)$ with $\Phi_i(\rho)=\Tr_B U(\rho \otimes \dyad{i})U^\dag$ when the randomness source $\sigma$ has the spectral decomposition of $\sigma=\sum_i p_i \dyad{i}$. We define the complementary channel for each $\Phi_i$ as $\Tilde{\Phi}_i(\rho)=\Tr_A U(\rho \otimes \dyad{i})U^\dag$. Note that $\Tilde{\Phi}:=\sum_i p_i \Tilde{\Phi}_i$ should be a constant channel from the definition of randomness-utilizing processes, thus $C_{EA}(\Tilde{\Phi})=0$.

Using the following expression \cite{bennett1999entanglement,bennett2002entanglement} of the entanglement-assisted classical capacity of $\mathcal{N}:A' \to B$,
\begin{equation}
    \max_{\phi_{AA'}} I(A:B)_{\tau_{AB}} = C_{EA}(\mathcal{N}),
\end{equation}
where $\phi_{AA'}$ is a pure state on $AA'$ and $\tau_{AB} = (\mathds{1}_A \otimes \mathcal{N}_{A' \to B})(\phi_{AA'})$, we get the following bound by applying Lemma \ref{lem:cap} for each $\Phi_i$ and $\Tilde{\Phi}_i$,
\begin{equation} \label{ineq2}
    \max\{I(R:A)_{\tau_{RA}}-C_{EA},I(R:B)_{\tau_{RB}}\} \leq -\log p_i,
\end{equation}
for an arbitrarily given bipartite pure state $\phi_{RA}$ with $\tau_{RAB}=(\mathds{1}_R \otimes U)(\phi_{RA}\otimes \dyad{i}_B)(\mathds{1}_R \otimes U^\dag)$ and $C_{EA}:=C_{EA}(\Phi)$. From the information conservation law for pure tripartite states \cite{lie2020randomness},
\begin{equation}
    2S(R)=I(R:A)+I(R:B),
\end{equation}

by choosing an arbitrary maximally entangled state $\phi_{RA}$ we get

\begin{equation}
    \max\{2\log d - C_{EA} -I,+I\} \leq -\log p_i,
\end{equation}
where $I:=I(R:B)_{\tau_{RB}}$. Now, for classical catalysis, $U$ should be a conditional unitary conditioning on the eigenbasis of $\sigma$, so we get $I=0$. The lower bound $S_{\min{}} (\sigma) = -\max_i \log p_i \geq 2\log d - C_{EA}$ follow from the minimization over $i$. The general bound for quantum catalysis follows from the minimization the lower bound over $I$, which is achieved at $I=\log d - \frac{1}{2}C_{EA}$, and we get $S_{\min{}}(\sigma)\geq \log d - \frac{1}{2}C_{EA}. $

\end{proof}

 For example, by noting that a dephasing map has $C_{EA}=\log d$ and the erasure map has $C_{EA}=0$, the known bounds for randomness costs for dephasing maps and erasure maps \cite{boykin2003optimal,boes2018catalytic, lie2020randomness} can be derived from Theorem \ref{thm:bound}. Note that Theorem \ref{thm:bound} shows the existence of a gap between classical and quantum bounds but the bounds may not be tight. For instance, there are some unital maps that do not permit classical catalytic implementation \cite{landau1993birkhoff}. Nevertheless, the min-entropy in the region between $\log d - \frac{1}{2} C_{EA}$ and $2\log d - C_{EA}$ is forbidden for any classical catalyst, we will say that catalysis with min-entropy in that region achieves the \textit{quantum advantage} of randomness usage. Hence, Theorem \ref{thm:nondeg} implies that the quantum advantage cannot be attained if the randomness source is non-degenerate.

 We summarize the implication of the previous results for the two most important randomness-utilizing process as the following corollary.

\begin{corollary}
    If the randomness source of a $d$-dimensional randomness-utilizing dephasing (erasure) map is non-degenerate, it should have the min-entropy larger than or equal to $\log d$ (\;$2\log d$\;).
\end{corollary}
\begin{figure}
    
    \includegraphics[width=.45\textwidth]{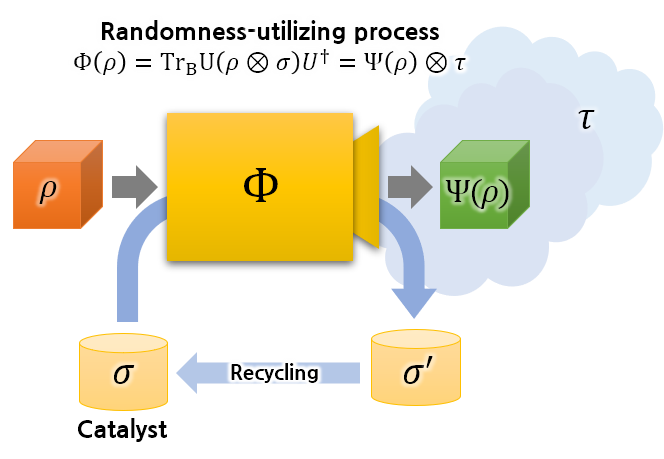}
    \centering
    \caption{A generalized randomness-utilizing process $\Phi$. If one intends to implement a certain quantum map $\Psi$ utilizing a randomness source $\sigma$ which has large enough min-entropy but is not a uniform random state, it could be implemented if one employs a broader notion of catalycity and allows the uncorrelated leftover randomness in the output state. }
    \label{fig:proc}

\end{figure}

This lower bound is twice larger than the minimal values of $\frac{1}{2}\log d$ for dephasing maps \cite{boes2018catalytic} and $\log d$ for erasure maps \cite{lie2019unconditionally, lie2020randomness}. Considering that the maximally mixed state, which could attain the minimal randomness cost, can be arbitrarily close to a non-degenerate state, we can see that being uniform is the key property for a quantum randomness source.

On the other hand, classical randomness source need not be uniform to function properly. For example, a non-degenerate randomness source given as $(1/8,3/8,1/2)$ can implement a dephasing map. See that by applying $I$ for the first and the second outcome and applying the Pauli $Z$ operator to a qubit system for the last outcome, one can completely dephase the qubit with respect to the computational basis. More generally, for a given probability distribution $\{p_m\}$, if one can find a family of real parameters $\{\theta_{nm}\}$ such that
\begin{equation}
    \sum_m p_m \exp i(\theta_{nm}-\theta_{n'm}) = \delta_{nn'},
\end{equation}
then one can dephase a quantum system with a randomness source with the spectrum $\{p_m\}$ and the set of unitary operators $\{Z_m:=\sum_n \exp(i \theta_{nm}) \dyad{n}\}$. However, to the best of our knowledge, there is no known complete characterization of classical randomness source that can be used for dephasing or erasure maps. The contrast against classical randomness characterizes uniformness as the essence of quantum catalytic randomness.

\subsection{Utilization of non-uniform randomness} \label{subsec:nonuni}

Are generic non-degenerate randomness sources useless as a quantum randomness source, after all? We show that, if we generalize the definition of randomness-utilizing process, any randomness source with high enough min-entropy can be used as a quantum randomness source. We will say that a quantum map $\Phi$ is a \textit{generalized randomness-utilizing} implementation of another process $\Psi$ on $\mathcal{B(H}_A)$ if there exists a bipartite unitary $U$ on $\mathcal{H}_A \otimes \mathcal{H}_B$ and a randomness source $\sigma$ such that
\begin{equation} \label{eqn:gen}
    \Phi(\rho)=\Tr_B U(\rho \otimes \sigma) U^\dag = \mathcal{T}(\Psi(\rho)),
\end{equation}
where $\mathcal{T}$ is an invertible quantum map, i.e. there exists another quantum map $\mathcal{R}$ such that $\mathcal{R} \circ \mathcal{T} =\mathcal{I}$.  This generalized definition says that, intuitively, if we can restore the output of the desired process deterministically from the output of an actually implemented process, we will consider it legitimate implementation. However, from the result of Ref. \cite{nayak2007invertible}, every invertible quantum map can be expressed as paring with an ancillary state followed by a unitary operation, i.e. the form of (\ref{eqn:rand}) without partial trace $\Tr_B$. Thus, by making $U$ in (\ref{eqn:gen}) absorb the unitary operators in $\mathcal{T}$, we can actually re-express the definition of generalized randomness-utilizing implementation $\Phi$ of process $\Psi$
\begin{equation} \label{eqn:gen2}
    \Phi(\rho)=\Tr_B U(\rho \otimes \sigma) U^\dag = \Psi(\rho)\otimes \tau,
\end{equation}
with some constant quantum state $\tau$ independent of input $\rho$. (See FIG. \ref{fig:proc}) In every practical sense, this definition is operationally legitimate. Every machine producing a certain type of product always produces accompanying byproducts such as noise, heat, dust or vibration. Nevertheless, as long as those byproducts can be unambiguously separated from the desired output, it is natural to say that the process was implemented as desired. Therefore we will call the uncorrelated byproduct $\tau$ of (\ref{eqn:gen2}) as the leftover randomness of the randomness-utilizing process $\Phi$.

We also generalize the notion of catalycity. If the residue randomness of $\Phi$ in (\ref{eqn:gen2}) can be repeatedly used for another generalized randomness-utilizing implementation (which can be different from the original implementation) of the same process as the randomness source, we will say that the randomness usage in the implementation is catalytic. This generalization is also operationally reasonable since the exact form of a catalyst need not be preserved as long as its `catalytic power' is conserved during the process. This generalization is depicted in FIG. \ref{fig:proc} as the transformation of the randomness source $\sigma$ to $\sigma'$, which can be recycled for another round of randomness-utilizing process.

We remark that in this generalized setting, non-decreasing property of randomness is \textit{not} forced unlike the original setting. The proof of Proposition \ref{prop:major} depends on the fact that the output system of the process has the same dimension as the input system, but in the generalized setting the output system can be much larger than the input system. In fact, extracting randomness of a randomness source and injecting it into the output state is allowed, therefore randomness can be actually \textit{consumed} in this setting. 

Nevertheless, in this generalized setting, it is indeed possible to catalytically use a non-degenerate state as a quantum randomness source. The following Theorem is proved in Ref. \cite{lie2020minent}, and we state it here for completeness.

\begin{prop} \cite{lie2020minent} \label{thm:circ}
    Any quantum state $\sigma$ with $S_\infty (\sigma) \geq \log d $ (or $S_\infty (\sigma) \geq 2\log d $) can be catalytically used as the randomness source for a generalized randomness-utilizing implementation of a $d$-dimensional dephasing map (or the erasure map).
\end{prop}
A sketch of proof is as follows: by the Birkhoff-von Neumann theorem \cite{birkhoff1946three, von1953certain}, every finite probability distribution with the min-entropy larger than or equal to $\log d$ can be expressed as a convex sum of uniform distribution with the supporter of size $d$. Therefore, by conditionally generating a randomness source, we can randomly choose one of those uniform distributions and extract it. This randomness can be generated by creating its purification and distributing it to two local systems. It is possible because the creation of entangled pure state can be done via unitary operation. By using the extracted uniform randomness, we can implement the desired process. As a result, both parties have some leftover randomness but it is allowed from the definition of the generalized randomness-utilizing processes. A detailed proof can be found in Ref. \cite{lie2020minent}.

Proposition \ref{thm:circ} shows that when extra work space is allowed, one can generate `bound' randomness by sharing an entangled state in the extra space that can be used for uniformizing a non-degenerate randomness source. This, in a sense, demonstrates the usage of `catalyst for catalyst'. This type of `expanding space to achieve uniformity' was also used in Ref. \cite{brandao2015second}.

\section{Conclusion} \label{subsec:conc}
We showed that when randomness is utilized to implement quantum maps, it is not expendable but inevitably reusable. It follows from a new no-go result on multipartite quantum secret sharing, we named the no-secret theorem. Especially, for dimension-preserving channels, randomness sources \textit{cannot} be used non-catalytically and in general every randomness-utilizing channel can be catalytically implemented if it is implemented twice at a time. We further found that the quantum advantage of randomness is common for arbitrary randomness-utilizing processes and it requires uniformness of the randomness source. Even if the source's entropic measures are arbitrarily high, it cannot be used as a quantum catalyst if it is non-degenerate. These two restrictions distinguish the resource theory of randomness from other types of quantum resource theories, but we also found that allowing expansion of dimension after randomness-utilizing process could circumvent both restrictions. It was done by showing that it is still possible to take advantage of catalytic quantum randomness in the generalized setting if the randomness source's min-entropy is high enough.

We remark that we focused on exact realizations of catalysis in contrast to Ref. \cite{muller2018correlating, brandao2015second} where the framework was generalized to approximate realizations but with the cost of having to prepare arbitrary many and arbitrarily large catalysts to achieve the desired level of accuracy. This work is more relevant to a realistic situation where the user has one given randomness source, not a set of multiple sources, and tries to assess its capability for various tasks. Furthermore, Theorem \ref{thm:bound} can be applied for arbitrary quantum maps, hence actually one can still use the results of this work to analyze approximate catalysis.

An interesting direction for future works is proving the existence of and constructing catalytic implementations achieving the lower bounds of Theorem \ref{thm:bound} for both classical and quantum catalyst cases. Another intriguing topic is rigorously establishing the resource theory of uncorrelatedness of randomness sources as mentioned in this work. Also it would be interesting to investigate the inclusion relation of FIG. \ref{fig:inc}. If it turns out that $RU=Cat$, then it would imply that quantum randomness has quantitative but no qualitative advantage compared to classical randomness. On the other hand, if $Cat=Unital$, then it would imply that there are some unital maps that must leak some information to whatever system it interacts with to implement the channel.

\begin{acknowledgments} 
This work was supported by the National Research Foundation of Korea (NRF) through grants funded by the the Ministry of Science and ICT (Grants No. NRF-2019M3E4A1080074 and No. NRF-2020R1A2C1008609).
\end{acknowledgments} 

\bibliography{main}

\end{document}